\documentclass[aps,pra,twocolumn,superscriptaddress]{revtex4-1}
\usepackage{graphicx}
\usepackage{dcolumn}
\usepackage{bm}
\usepackage{multirow}
\usepackage{ulem}
\usepackage{color}
\usepackage{subfigure}
\usepackage{amsmath}
\usepackage{upgreek}
\usepackage{amsthm,amsmath,amssymb}

\begin{document}
\title{Determination of hyperfine splittings and Land\'{e} $g_J$ factors of $5s~^2S_{1/2}$ and $5p~^2P_{1/2,3/2}$ states of $^{111,113}$Cd$^+$ for a microwave frequency standard}

\author{J. Z. Han}
\email{These authors contributed equally to this work.}
\affiliation{
State Key Laboratory of Precision Measurement Technology and Instruments, Key Laboratory of Photon Measurement and Control Technology of Ministry of Education, Department of Precision Instrument, Tsinghua University, Beijing 100084, China}%
\author{R. Si}
\email{These authors contributed equally to this work.}
\affiliation{
Shanghai EBIT Lab, Key Laboratory of Nuclear Physics and Ion-beam Application, Institute of Modern Physics, Department of Nuclear Science and Technology, Fudan University, Shanghai 200433, China}%
\author{H. R. Qin}
\affiliation{
Department of Physics, Tsinghua University, Beijing 100084, China}%
\author{N. C. Xin}
\affiliation{
State Key Laboratory of Precision Measurement Technology and Instruments, Key Laboratory of Photon Measurement and Control Technology of Ministry of Education, Department of Precision Instrument, Tsinghua University, Beijing 100084, China}%
\author{Y. T. Chen}
\affiliation{
State Key Laboratory of Precision Measurement Technology and Instruments, Key Laboratory of Photon Measurement and Control Technology of Ministry of Education, Department of Precision Instrument, Tsinghua University, Beijing 100084, China}%
\author{S. N. Miao}
\affiliation{
State Key Laboratory of Precision Measurement Technology and Instruments, Key Laboratory of Photon Measurement and Control Technology of Ministry of Education, Department of Precision Instrument, Tsinghua University, Beijing 100084, China}%
\author{C. Y. Chen}
\email{chychen@fudan.edu.cn}
\affiliation{
Shanghai EBIT Lab, Key Laboratory of Nuclear Physics and Ion-beam Application, Institute of Modern Physics, Department of Nuclear Science and Technology, Fudan University, Shanghai 200433, China}%
\author{J. W. Zhang}
\email{zhangjw@tsinghua.edu.cn}
\affiliation{
State Key Laboratory of Precision Measurement Technology and Instruments, Key Laboratory of Photon Measurement and Control Technology of Ministry of Education, Department of Precision Instrument, Tsinghua University, Beijing 100084, China}%
\author{L. J. Wang}
\email{lwan@mail.tsinghua.edu.cn}
\affiliation{
State Key Laboratory of Precision Measurement Technology and Instruments, Key Laboratory of Photon Measurement and Control Technology of Ministry of Education, Department of Precision Instrument, Tsinghua University, Beijing 100084, China}%
\affiliation{
Department of Physics, Tsinghua University, Beijing 100084, China}%

\begin{abstract}

Regarding trapped-ion microwave-frequency standards, we report on the determination of hyperfine splittings and Land\'{e} $g_J$ factors of $^{111,113}$Cd$^+$. The hyperfine splittings of the $5p~^2P_{3/2}$ state of $^{111,113}$Cd$^+$ ions were measured using laser-induced fluorescence spectroscopy. The Cd$^+$ ions were confined in a linear Paul trap and sympathetically cooled by Ca$^+$ ions. Furthermore, the hyperfine splittings and Land\'{e} $g_J$ factors of the $5s~^2S_{1/2}$ and $5p~^2P_{1/2,3/2}$ levels of $^{111,113}$Cd$^+$ were calculated with greater accuracy using the multiconfiguration Dirac--Hartree--Fock scheme. The measured hyperfine splittings and the Dirac--Hartree--Fock calculation values were cross-checked, thereby further guaranteeing the reliability of our results. The results provided in this work can improve the signal-to-noise ratio of the clock transition and the accuracy of the second-order Zeeman shift correction, and subsequently the stability and accuracy of the microwave frequency standard based on trapped Cd$^+$ ions.

\end{abstract}
\date{\today}

\maketitle
\section{Introduction}
With their improvements in accuracy over time, atomic clocks have played an important role in practical applications \cite{hinkley2013atomic, burt2021demonstration} and testing the fundamental physics \cite{dzuba2016strongly, wcislo2016experimental, safronova2018search}. Indeed, the microwave-frequency atomic clock plays a vital role in satellite navigation \cite{bandi2011high}, deep space exploration \cite{prestage2007atomic, burt2016jpl}, and timekeeping \cite{diddams2004standards}. Among the many clock proposals, trapped-ion microwave-frequency clocks have attracted wide attention from researchers because the ions are well isolated from the external environment in an ion trap. The setup is conducive to improvements in the transportability of atomic clocks\cite{schwindt2016highly,mulholland2019laser, mulholland2019compact, hoang2021integrated}. Such clocks are also considered the next generation of practical microwave clocks \cite{schmittberger2020review}.

Cadmium ions (Cd$^+$) benefit from a simple and distinct electronic structure, which is easily controlled, manipulated, and measurable with high precision. The microwave-frequency standard based on laser-cooled $^{113}$Cd$^+$ has achieved an accuracy of $1.8\times10^{-14}$ and a short-term stability of $4.2\times10^{-13}/\sqrt{\tau}$ \cite{miao2021precision}. The high performance and potential for miniaturization make this frequency standard suitable in establishing a ground-based transportable frequency reference for navigation systems and for comparing atomic clocks between remote sites \cite{zhang2012, WangS.2013, Miao2015, miao2021precision}. Moreover, it has been proposed as a means to achieve an ultra-high level of accuracy down to $10^{-15}$ \cite{han2021toward}, highlighting the importance of accurately evaluating systematic frequency shifts.

Optical pumping is a fundamental process in operating a trapped-ion microwave frequency standard. The optical pumping efficiency determines directly the signal-to-noise ratio of the ``clock signal," which affects the short-term stability and measurement accuracy of the ground-state hyperfine splitting (HFS) for such frequency standards. Realizing optical pumping for the $^{113}$Cd$^+$ microwave-frequency standard requires a blueshift in the laser frequency of the Doppler-cooling transition $5s~^2S_{1/2}~F=1,~m_F=1$--$5p~^2P_{3/2}~F=2,~m_F=2$ to reach the $5p~^2P_{3/2}~F=1$ hyperfine level. However, there are no precise measurements available of the HFSs for other excited states \cite{li2018relativistic}. A preliminary measurement of the HFS for the $5p~^2P_{3/2}$ level of the $^{113}$Cd$^+$ ion is approximately 800~MHz \cite{tanaka1996determination}.
Therefore, to improve the optical pumping efficiency and hence the performance of the $^{113}$Cd$^+$ microwave frequency standard, the HFSs of the $5p~^2P_{3/2}$ level of the $^{113}$Cd$^+$ ion need to be determined with greater accuracy. From the perspective of atomic structure calculations, the high-precision measurements of the HFSs for the $5p~^2P_{3/2}$ level of $^{111,113}$Cd$^+$ can also be used for testing and developing calculation models of the atomic structure.

In a trapped-ion microwave frequency standard, an external magnetic field is applied to provide the quantization axis to break the degeneracy of the ground-state magnetic level. Among all the systematic frequency shifts of a frequency standard, one dominant shift is the second-order Zeeman shift (SOZS) induced by the external magnetic field \cite{berkeland1998laser, phoonthong2014determination, miao2021precision}. The precise estimation of this SOZS and the calibration of the external magnetic field require accurate knowledge of the ground-state Land\'{e} $g_j$ factor \cite{han2019roles}. The external magnetic field in our latest laser-cooled microwave-frequency standard based on trapped $^{113}$Cd$^+$ ions is approximately $8000$ nT \cite{miao2021precision}. However, only two theoretical studies have provided a value of the ground-state Land\'{e} $g_J$ factor of Cd$^+$, one giving $2.00286(53)$, calculated using the relativistic-coupled-cluster (RCC) theory \cite{han2019roles}, and the other giving $2.002291(4)$, calculated by the $\Lambda$-approach RCC ($\Lambda$-RCC) theory \cite{yu2020ground}. The two Land\'{e} $g_J$ factors have a difference of $0.0006$ that generates a relative frequency shift of $6.6\times10^{-14}$. This large systematic shift obviously falls short inaccuracy of our latest $^{113}$Cd$^+$ microwave-frequency standard ($1.8\times10^{-14}$) \cite{miao2021precision}. Therefore, re-determining the ground-state $g_J$ factor of $^{113}$Cd$^+$ is imperative if further improvements in accuracy for this microwave-frequency standard are to be attained.

In this work, the HFSs of the $5p~^2P_{3/2}$ level of the $^{113}$Cd$^+$ ion is measured using the laser-induced fluorescence (LIF) technique. To maintain a low-temperature environment, the $^{113}$Cd$^+$ ions are sympathetic-cooled by laser-cooled $^{40}$Ca$^+$ ions, a technique that improves the accuracy of measurements. Furthermore, the HFSs and Land\'{e} $g_J$ factors of both the $5s~^2S_{1/2}$ and $5p~^2P_{1/2,3/2}$ levels were calculated using the multiconfiguration Dirac--Hartree--Fock (MCDHF) method. Electron correlation effects are carefully investigated and taken into account. Off-diagonal terms are also included to improve the calculation accuracy of the HFSs for the $5p~^2P_{1/2,3/2}$ level in Cd$^+$. Cross-checking the measured and calculated HFS results ensures the reliability and accuracy of the results provided in this work.
Our results are of great importance for further improving the performance of the Cd$^+$ microwave-frequency standard.

\section{Experiment}
To obtain the HFSs of the $5p~^2P_{3/2}$ level for $^{111,113}$Cd$^+$, we first measure the frequency shifts from the $5s~^2S_{1/2}~F=1$--$5p~^2P_{3/2}~F=2$ transition of $^{111,113}$Cd$^+$ to the $5s~^2S_{1/2}$--$5p~^2P_{3/2}$ transition of $^{114}$Cd$^+$. Briefly, for the experimental setup (see Ref.~\cite{han2021toward} for details), we prepared crystals of two ion species consisting of approximately $10^5$ Ca$^+$ and Cd$^+$ ions in a linear Paul trap. The Ca$^+$ and Cd$^+$ ions are produced by selected-photoionization using laser beams of wavelength 423-nm (Ca $4s^2~^1S_0$--$4s4p~^1P_1$) / 374-nm (Ca $4s4p~^1P_1$-- Continuum), and 228-nm (Cd $5s^2~^1S_0$--$5s5p~^1P_1$). The Ca$^+$ are used as coolant ions that are Doppler-cooled using lasers beams of wavelength 397-nm (Ca$^+$ $4s~^2S_{1/2}$--$4p~^2P_{1/2}$) and 866-nm (Ca$^+$ $3d~^2D_{3/2}$--$4p~^2P_{1/2}$). The Cd$^+$ ions are sympathetically-cooled to less than 0.5 K through Coulomb interactions with the Ca$^+$ ions. The frequency shifts of the $5s~^2S_{1/2}~(F=1)$--$5p~^2P_{3/2}~(F=2)$ transition of $^{111,113}$Cd$^+$ and the $5s~^2S_{1/2}$--$5p~^2P_{3/2}$ transitions of $^{114}$Cd$^+$ were measured using scanning frequencies in a weak 214.5-nm probe laser beam. The $5s~^2S_{1/2}~(F=1,m_F=1)$--$5p~^2P_{3/2}~(F=2,m_F=2)$ transition is a cycling transition that was used to cool and detect the $^{111,113}$Cd$^+$ ions. Although the circularly polarized cooling laser beam excites a cycling transition, ions may, as a result of the polarization impurity, still, leak to the $5s~^2S_{1/2}~F=0$ state via $5p~^2P_{3/2}~F=1$ state. To increase detection efficiency, 20-dBm microwave radiation resonant with the ground-state hyperfine transition (15.2-GHz for $^{113}$Cd$^+$ and 14.5-GHz for $^{111}$Cd$^+$) is applied during LIF detection. The frequency of each laser beam is measured using a high-precision wavemeter (HighFinesse WS8-2).

\begin{figure*}
\centering
\resizebox{0.8\textwidth}{!}{
\includegraphics{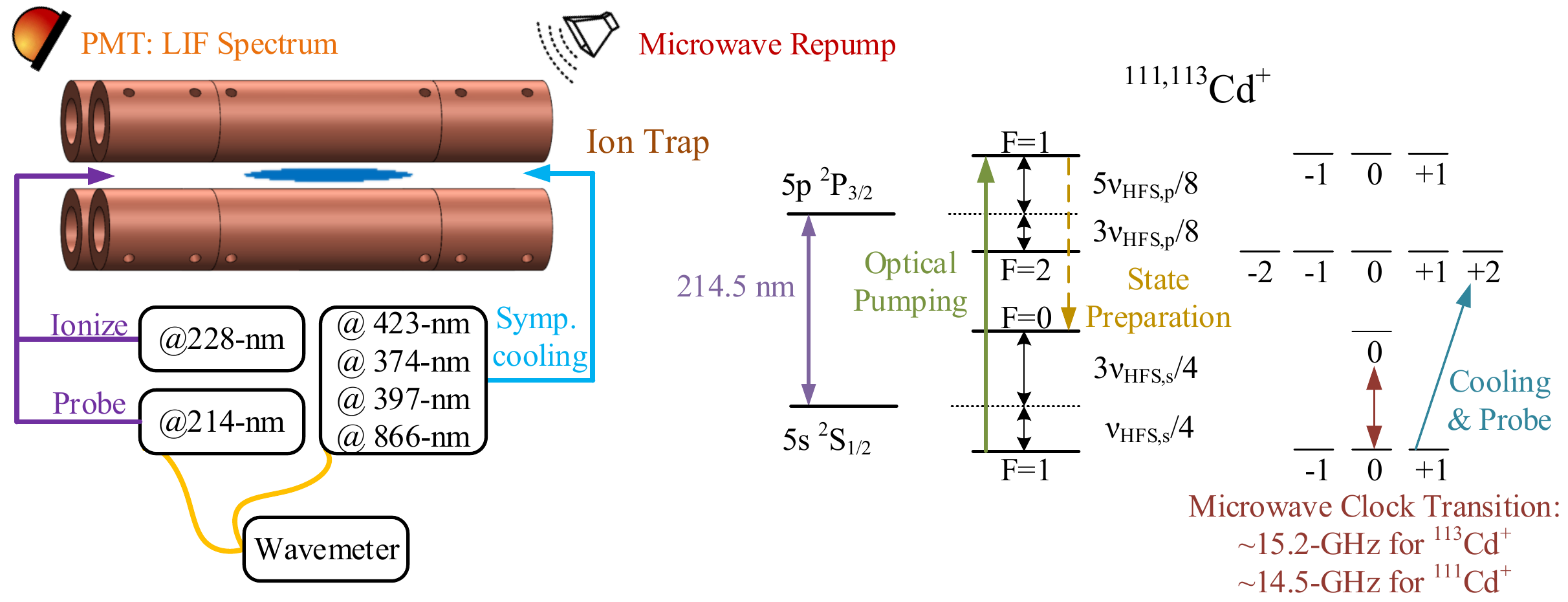}
}
\caption{ Schematic of the experiment setup for the $^{111,113}$Cd$^+$ HFS measurements. Sympathetic cooling technology is used to maintain the Cd$^+$ cloud at low temperatures and improve measurement accuracy. The energy level scheme (not to scale) for $^{111,113}$Cd$^+$ is also presented.}
\label{pic1}
\end{figure*}

\begin{figure}
\centering
\resizebox{0.40\textwidth}{!}{
\includegraphics{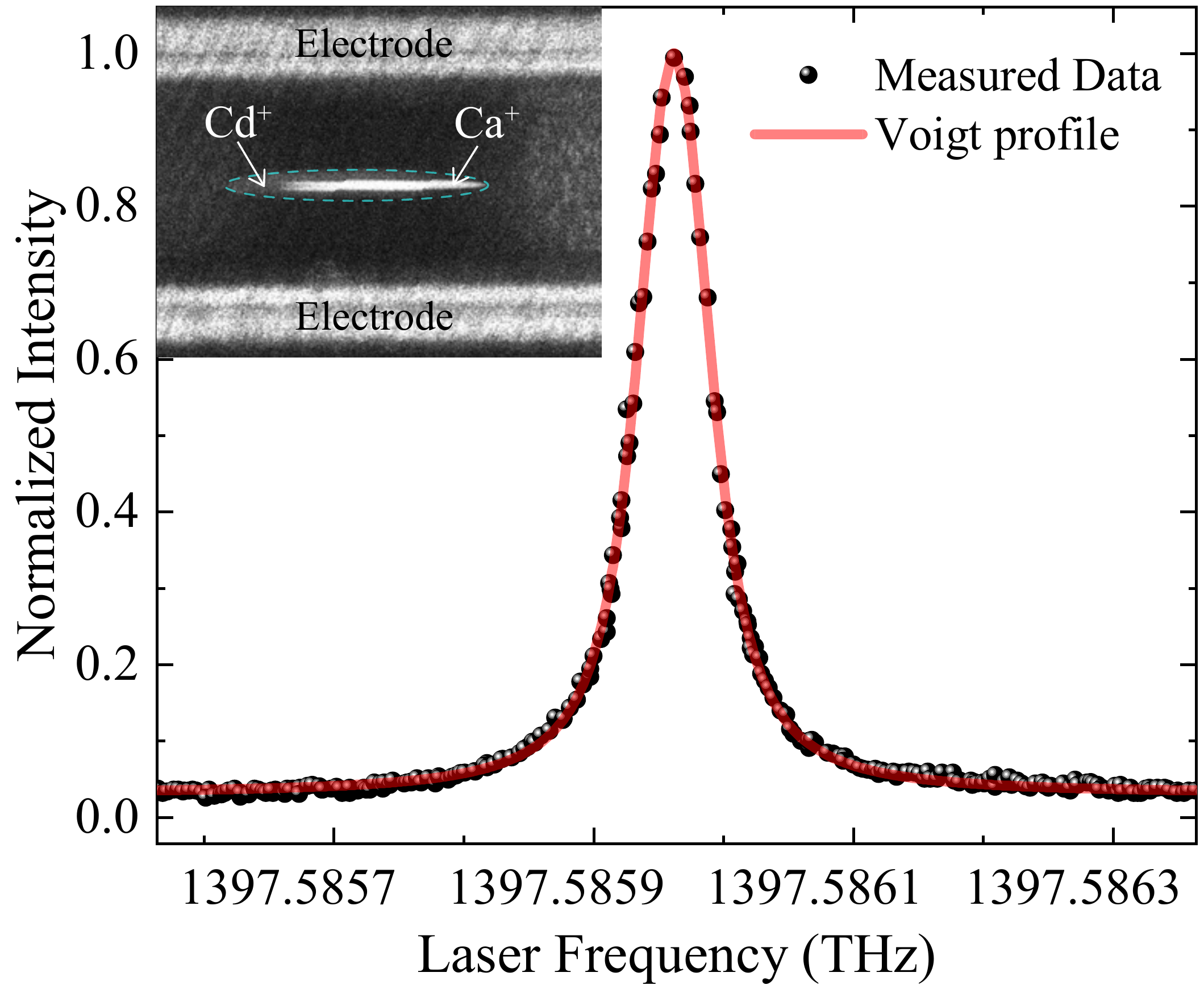}
}
\caption{Typical LIF spectrum of Cd$^+$, using $^{113}$Cd$^+$ as an example. The frequency of the probe laser beam (214 nm wavelength) is scanned around the resonance frequency over a range of 600~MHz. The measured line profiles are fitted with a Voigt function. The inset is an image of the two-species ion cloud of Cd$^+$ and Ca$^+$ captured by an EMCCD camera; aberration has blurred the image.}
\label{LIF}
\end{figure}

In obtaining the measured LIF spectrum (Fig.~\ref{LIF}), the beam intensity is maintained below 5 $\mu$W/mm$^2$ (saturation parameter 0.0006) to reduce the cooling and heating effects of the probe beam. The fitted curve is the Voigt profile \cite{zuo2019direct, han2021toward}, expressed as
\begin{eqnarray}
F&=&F_0+(F_L\ast F_G)(\nu),\nonumber \\
F_L(\nu)&=&\frac{2A}{\pi}
\frac{\nu_L}{4(\nu-\nu_c)^2+\nu_L^2},\nonumber \\
F_G(\nu)&=&\sqrt{\frac{4\ln2}{\pi}}
\frac{e^{-\frac{4\ln2}{\nu^2_G}\nu^2}}{\nu_G},
\end{eqnarray}
where $F_0$ is the offset, $\nu$ is the laser beam frequency, $\nu_c$ is the ion resonance frequency, $A$ is the area, $\nu_L$ is the Lorentzian width, $\nu_G$ is the Gaussian width of Doppler broadening. The line profile is slightly asymmetrically because of the heating and cooling effects of the probe beam, which lead to a slight increase in the uncertainty of the estimated transition frequency.

Measurements present three sources of uncertainty:
\begin{itemize}
\item[i)] Statistical uncertainties. For the Cd$^+$ $5s~^2S_{1/2}$--$5p~^2P_{3/2}$ transition, $\nu_L$ is 60.13~MHz which represent the natural width, the fitted $\nu_G$ is approximately 30~MHz, and the ion temperature is estimated to be approximately 100 mK. The statistical uncertainty associated with the transition frequencies of $^{114}$Cd$^+$ and the $^{111,113}$Cd$^+$ is approximately 1~MHz, and thus the statistical uncertainty in their differences is approximately 1.4~MHz;

\item[ii)] Instrument uncertainties. The uncertainty arising from the drift in the wavemeter is less than 0.5~MHz in a laboratory environment \cite{liu2018ultraviolet};

\item[iii)] Systematic uncertainties. Most systematic shifts are common to the $5s~^2S_{1/2}$--$5p~^2P_{3/2} $ transitions of both $^{114}$Cd$^+$ and $^{111/113}$Cd$^+$ and thus cancel each other out. Because the $5s~^2S_{1/2}~(F=1)$--$5p~^2P_{3/2}~(F=2)$ transition of Cd$^+$ is sensitive to magnetic fields, the Zeeman shift becomes the dominant contributor to systematic uncertainties.
\end{itemize}
In a weak field ($\mu_B B\sim0.14$~MHz $\ll$ hyperfine constant $A$), the Zeeman shift for a specific energy level is expressed as
\begin{equation}
E_{Zeeman}=g_F M_F \mu_B \cdot B,
\end{equation}
where $g_F$ is given by
\begin{equation}
g_F=\frac{F(F+1)+J(J+1)-I(I+1)}{2F(F+1)}g_J,
\end{equation}
where $J = L+S$ the total electron angular momentum ($S$ and $L$ the spin and orbital angular momenta), and $F=I+J$ the total angular momentum with $I$ denoting the nuclear spin. In typical conditions of our experiment, $B\sim8000$ nT \cite{Miao2015, miao2021precision}. By introducing values $g_J=2.002257$ and $1.334056$ for the levels $5s~^2S_{1/2}$ and $5p~^2P_{3/2}$ calculated in this work (see text below), the Zeeman shift for the $5s~^2S_{1/2}~(F=1)$--$5p~^2P_{3/2}~(F=2)$ transition of Cd$^+$ is estimated to be 0.11~MHz. Therefore, the total systematic shifts for the $5s~^2S_{1/2}~(F=1)$--$5p~^2P_{3/2}~(F=2)$ transitions of Cd$^+$ are estimated to be below 0.5~MHz.

The final frequencies for the $5s~^2S_{1/2}~(F=1)$--$5p~^2P_{3/2}~(F=2)$ transitions of $^{111,113}$Cd$^+$ and that for $5s~^2S_{1/2}$--$5p~^2P_{3/2}$ of $^{114}$Cd$^+$ are determined to be 4649.0(1.6)~MHz and 4041.8(1.6)~MHz, respectively.

In LS-coupling, the energy shifts after the hyperfine interaction are expressed as \cite{foot2005atomic}
\begin{equation}
E_{HFS}=A\langle I\cdot J \rangle = \frac{A}{2}[F(F+1)-I(I+1)-J(J+1)].
\end{equation}
Therefore, for $^{111,113}$Cd$^+$, we have
\begin{equation}
\nu_{s,F=1\rightarrow p,F=2}^{111/113}=\nu_{s\rightarrow p}^{111/113}
+\frac{1}{4}\nu_{HFS,s}-\frac{3}{8}\nu_{HFS,p},
\label{eq:hfs}
\end{equation}
where $\nu_{s,F=1\rightarrow p,F=2}^{111/113}$ is the transition frequency of $5s~^2S_{1/2}~(F=1)$--$5p~^2P_{3/2}~(F=2)$ of $^{111,113}$Cd$^+$; $\nu_{s\rightarrow p}^{111/113}$ is the transition frequency of $5s~^2S_{1/2}$--$5p~^2P_{3/2}$; $\nu_{HFS,s}$ is the HFS of $5s~^2S_{1/2}$; and $\nu_{HFS,p}$ is the HFS of $5p~^2P_{3/2}$. In reference to $^{114}$Cd$^+$, through a linear transformation, Eq. (\ref{eq:hfs}) may be expressed as
\begin{eqnarray}
\Delta \nu_{s,F=1\rightarrow p,F=2}^{111/113,114}
&=&\Delta \nu_{s\rightarrow p}^{111/113,114}
+\frac{1}{4}\nu_{HFS,s}^{111/113}\\\nonumber
&-&\frac{3}{8}\nu_{HFS,p}^{111/113},
\end{eqnarray}
where $\Delta \nu_{s,F=1\rightarrow p,F=2}^{111/113,114}=\nu_{s,F=1\rightarrow p,F=2}^{111/113}-\nu_{s\rightarrow p}^{114}$ and $\Delta \nu_{s\rightarrow p}^{111/113,114}=\nu_{s\rightarrow p}^{111/113}-\nu_{s\rightarrow p}^{114}$. With our measurements, $\Delta \nu_{s,F=1\rightarrow p,F=2}^{111/113,114}$ are respectively 4649.0(1.6)~MHz and 4041.8(1.6)~MHz, whereas $\Delta \nu_{s\rightarrow p}^{111/113,114}$ are 1314.3(22)[023]~MHz and 555.2(23)[008]~MHz \cite{hammen2018calcium}. From our previous measurements obtained through double-resonance microwave laser spectroscopy, the $\nu_{HFS,s}^{111/113}$ were accurately measured to be 14530507349.9(1.1) Hz \cite{zhang2012} and 15199862855.02799(27) Hz \cite{miao2021precision}, from which we derived $\nu_{HFS,p}^{111/113}$ to be 794.6(3.6)~MHz and 835.5(2.9)~MHz.

\section{Theory}
\subsection{Multiconfiguration Dirac--Hartree--Fock approach}
The MCDHF method~\cite{Fischer2016}, as implemented in the {\sc Grasp} computer package~\cite{Jonsson2013,Fischer2019}, is employed to obtain wave functions referred to as atomic state functions. Specifically, they are approximate eigenfunctions of the Dirac Hamiltonian describing a Coulombic system given by
\begin{equation}
H_{\rm DC}= \sum_{i=1}^N(c~\bm{\alpha_i}\cdot\bm{p_i}
+(\beta_i-1)c^2+V_i)+\sum_{i<j}^N\frac{1}{r_{ij}},
\end{equation}
where $V_i$ denotes the monopole part of the electron--nucleus interaction for a finite nucleus and $r_{ij}$ the distance between electrons $i$ and $j$; $\bm{\alpha}_i$ and $\beta_i$ are the Dirac matrices for electron $i$.

Electron correlations are included by expanding $|\Gamma J \rangle$, an atomic state function, over a linear combination of configuration state functions (CSFs) $|\gamma J\rangle$,
\begin{equation}
|\Gamma J \rangle =\sum_{\gamma} c_{\gamma} | \gamma J\rangle,
\end{equation}
where $\gamma$ represents the parity and all the coupling tree quantum numbers needed to define the CSF uniquely. The CSFs are four-component spin-angular coupled, antisymmetric products of Dirac orbitals of the form
\begin{equation}
\phi({\bf r})=\frac{1}{r}\left(\begin{array}{c}
P_{n\kappa}(r)\chi_{\kappa m }(\theta,\phi)\\
iQ_{n\kappa}(r)\chi_{-\kappa m}(\theta,\phi)\end{array}\right).
\end{equation}
The radial parts of the one-electron orbitals and the expansion coefficients $c_{\gamma}$ of the CSFs are obtained by the self-consistent relativistic field (RSCF) procedure. In the following calculations of the relativistic configuration interaction (RCI), the Dirac orbitals are kept fixed, and only the expansion coefficients of the CSFs are determined for selected eigenvalues and eigenvectors of the complete interaction matrix. This procedure includes the Breit interaction and the leading quantum electrodynamic (QED) effects (vacuum polarization and self-energy).

The restricted active-set method is used in obtaining the CSF expansions by allowing single and double (SD) substitutions from a selected set of reference configurations to an active set (AS) of given orbitals. The configuration space is increased step by step by increasing the number of layers, specifically, a set of virtual orbitals. These virtual orbitals are optimized in the RSCF procedure while all orbitals of the inner layers are fixed.

The interaction between the electrons and the electromagnetic multipole moments of the nucleus splits each fine structure level into multiple hyperfine levels. The interaction couples the nuclear spin $\bm{I}$ with the total electronic angular momentum $\bm{J}$ to obtain total angular momentum $\bm{F=I+J}$.

The hyperfine contribution to the Hamiltonian is represented by a multipole expansion
\begin{equation}
H_{\rm HFS}=\sum_{k\geq1}\bf{T^{(k)}}\cdot \bf{M^{(k)}},
\end{equation}
where $\bf{T^{(k)}}$ and $\bf{M^{(k)}}$ are spherical tensor operators of rank $k$ in the electronic and nuclear spaces, respectively~\cite{LINDGREN1975}. The $k=1$ term represents the magnetic dipole interaction, and the $k=2$ term the electric quadrupole interaction. Higher-order terms are minor and can often be neglected.

For the scheme considered in this work ($^{111,113}$Cd$^+$ with $I=1/2$), only the magnetic dipole interaction is non-zero. To first-order, the fine-structure level $\gamma J$ is then split according to
\begin{equation}
\begin{aligned}
& \langle \Gamma IJFM_F| \bm{T^{(1)}\cdot M^{(1)}} | \Gamma IJFM_F \rangle \\
= & (-1)^{I+J+F}
\left\{ \begin{array}{ccc}
I & J & F\\
J & I & 1
\end{array}\right\}
\langle \Gamma J || T^{(1)} || \Gamma J \rangle
\langle \Gamma I || M^{(1)} || \Gamma I \rangle,
\end{aligned}
\end{equation}
where the coefficient in curly brackets in the 6j symbol of the rotation group. The reduced matrix elements of the nuclear tensor operators are related to the conventional nuclear magnetic dipole moment,
\begin{equation}
\langle \Gamma I || M^{(1)} || \Gamma I \rangle
= \mu_I \sqrt{\frac{(2I+1)(I+1)}{I}}.
\end{equation}
The hyperfine interaction energy contribution to a specified hyperfine level is then given by
\begin{equation}
\langle \Gamma IJFM_F| \bm{T^{(1)}\cdot M^{(1)}} | \Gamma IJFM_F \rangle
= \frac{1}{2}A_JC,
\end{equation}
where $A_J$ is the magnetic dipole hyperfine constants
\begin{equation}
A_J=\frac{\mu_I}{I}\frac{1}{\sqrt{J(J+1)}}
\langle \gamma J || T^{(1)} || \gamma J \rangle ,
\end{equation}
and $C=F(F+1)-J(J+1)-I(I+1)$.

However, this method discards the off-diagonal hyperfine interaction. This is not sufficient for $^2P_{1/2,3/2}$ because the hyperfine interaction between the two $F=1$ hyperfine levels is non-negligible owing to their minor splitting. To account for the off-diagonal hyperfine effects, we consider the second-order hyperfine interaction between $^2P_{1/2,3/2}$. The contribution associated with sublevel labelled $\gamma IJFM_F$ may be expressed as
\begin{equation}
\frac{|\langle \gamma IJFM_F |H_{hfs}| \gamma'IJ'FM_F \rangle|^2}{E_J-E_{J'}}.
\end{equation}

In the relativistic theory, choosing the direction of the magnetic field as the $z$-direction of the interaction and neglecting all diamagnetic contributions, the interaction between the magnetic moment of the atom and an external field may be written as
\begin{equation}
H_M=(N_0^{(1)}+\Delta N_0^{1})B,
\end{equation}
where the last term is the Schwinger QED correction~\cite{Cheng1985}. To first order, the fine-structure splitting in the energy level is
\begin{equation}
\begin{aligned}
& \langle \Gamma JM_J | N_0^{(1)}+\Delta N_0^{1} | \Gamma JM_J \rangle B \\
= & \frac{M_J}{\sqrt{J(J+1)}} \langle \Gamma J || \bm{N^{(1)}+\Delta N^{(1)}} || \Gamma J \rangle B.
\end{aligned}
\end{equation}
Factoring out the dependence on the $M_J$ quantum number, the energies are expressed in terms of the Land\'{e} $g_J$ factor
\begin{equation}
g_J=2\frac{\langle \Gamma J || \bm{N^{(1)}+\Delta N^{(1)}} || \Gamma J\rangle}
{\sqrt{J(J+1)}}.
\end{equation}
The energy splittings are then given by
\begin{equation}
g_JM_J\frac{B}{2}.
\end{equation}

\begin{table}
\setlength{\tabcolsep}{1pt}
\footnotesize
\centering
\caption{Calculated hyperfine splitting (HFS) (in~MHz) and Land\'{e} $g_J$ factors for $4d^{10}5s\ ^2S_{1/2}$ of $^{111,113}$Cd$^+$ obtained through the MCDHF approach.}\label{tab_S1}
\begin{tabular}{llllllllllll}
\hline
\hline
& \multicolumn{3}{c}{MCDHF2+RCI} & \multicolumn{3}{c}{MCDHF3+RCI}\\
& $^{111}$HFS & $^{113}$HFS & $g_J$ & $^{111}$HFS & $^{113}$HFS & $g_J$\\
\hline
5 & 11841 & 12386 & 2.002243 & 12925 & 13521 & 2.002242 \\
6 & 13421 & 14040 & 2.002245 & 14096 & 14746 & 2.002247 \\
7 & 13671 & 14301 & 2.002254 & 14414 & 15079 & 2.002250 \\
8 & 13879 & 14518 & 2.002260 & 14476 & 15143 & 2.002256 \\
9 & 13884 & 14524 & 2.002262 & 14515 & 15184 & 2.002257 \\
10 & 13925 & 14567 & 2.002263 & 14541 & 15211 & 2.002257 \\
11 & 13919 & 14561 & 2.002263 & 14532 & 15202 & 2.002257 \\
12 & 13921 & 14562 & 2.002262 & 14537 & 15207 & 2.002257 \\
Final & 13921(7) & 14563(7) & 2.002262(2) & 14536(9) & 15206(9) & 2.002257(1)\\
\hline
\end{tabular}
\end{table}

\subsection{Computation Strategy}
\begin{figure}
\includegraphics[width=0.95\columnwidth]{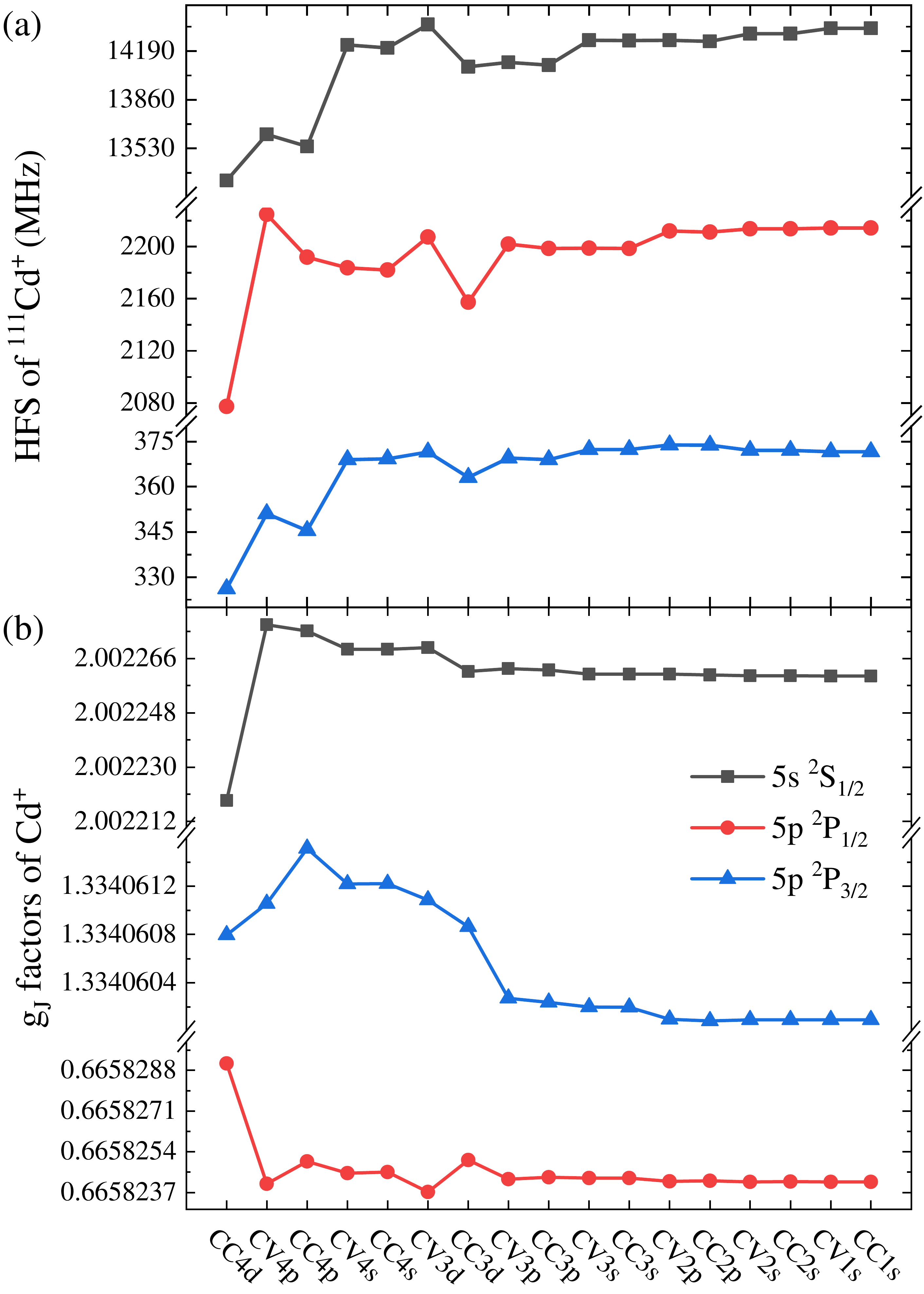}
\caption{Hyperfine structure constants $A_J$ and Land\'{e} $g_J$ factors of the $4d^{10}5s\ ^2S_{1/2}$ and $4d^{10}5p\ ^2P_{1/2,3/2}$ levels for $^{113}$Cd$^+$, from the various correlation models used in our MCDHF1+RCI calculation.}
\label{fig_CVCC}
\end{figure}

Initially, the MCDHF calculation was performed using the extended optimal-level scheme for the states of the $4d^{10}5s$ and $4d^{10}5p$ configurations, and these occupied orbitals were determined simultaneously and maintained throughout subsequent calculations. Because the $4f$ subshell in both the $4d^{10}5s$ and $4d^{10}5p$ configurations is vacant, imaging the strong pair correlations between $4d^2$ and $4f^2$, and between $4d^2$ and $5d^2$ is easy. The strong core-core (CC) correlation of the $4d$ electrons and the single valence $5s/5p$ electron do not allow us to include only the valence correlation; hence, our MCDHF calculation starts from the CC$_{4d}$ mode, in which the $5s/5p$ and $4d$ electrons can be SD-excited to the $n\leq8, l\leq6$ level (AS$_8$, where `8' labels the maximum principal quantum number in the corresponding AS). To investigate the correlation effects of the inner core electrons, we performed a few RCI calculations using the MCDHF wavefunctions derived from the CC$_{4d}$ calculation. This calculation method is labelled MCDHF1+RCI in this paper. From the plots of the $A_J$ and $g_J$ factors for the $^2S_{1/2}$ and $^2P_{1/2,3/2}$ levels of $^{113}$Cd$^{+}$ from this calculation method (Fig.~\ref{fig_CVCC}), we see that they are sensitive to the core correlations, and even the core--valence (CV) correlation of the $1s$ orbital. To describe the wavefunctions better, we performed a second calculation that includes the strong CV and CC correlations in the RSCF procedure rather than only including the CV/CC correlations in the RCI procedure.

\begin{table*}
\setlength{\tabcolsep}{4pt}
\footnotesize
\centering
\caption{Hyperfine splitting (HFS) (in~MHz) and Land\'{e} $g_J$ factors for the $4d^{10}5p\ ^2P_{1/2,3/2}$ level of $^{111,113}$Cd$^+$. HFS$_{\rm{n.o}}$ indicate that no off-diagonal contributions were included, whereas HFS$_{\rm{w.o}}$ refers to HFS calculations with off-diagonal contributions.}\label{tab_P1}
\begin{tabular}{llllllllll}
\hline
\hline
& \multicolumn{3}{c}{MCDHF2+RCI} & \multicolumn{5}{c}{MCDHF3+RCI}\\
& $^{111}$HFS$_{\rm{n.o}}$ & $^{113}$HFS$_{\rm{n.o}}$ & $g_J$ & $^{111}$HFS$_{\rm{n.o}}$ & $^{113}$HFS$_{\rm{n.o}}$ & $^{111}$HFS$_{\rm{w.o}}$ & $^{113}$HFS$_{\rm{w.o}}$ & $g_J$\\
\hline
\multicolumn{9}{c}{$4d^{10}5p\ ^2P_{1/2}$}\\
5 & 1816 & 1900 & 0.665833 & 2061 & 2156 & 2067 & 2150 & 0.665829 \\
6 & 2099 & 2196 & 0.665825 & 2267 & 2371 & 2270 & 2368 & 0.665821 \\
7 & 2054 & 2148 & 0.665825 & 2255 & 2359 & 2259 & 2355 & 0.665820 \\
8 & 2142 & 2241 & 0.665820 & 2310 & 2416 & 2314 & 2413 & 0.665814 \\
9 & 2123 & 2220 & 0.665822 & 2300 & 2406 & 2304 & 2403 & 0.665818 \\
10 & 2146 & 2245 & 0.665821 & 2324 & 2432 & 2328 & 2428 & 0.665818 \\
11 & 2144 & 2243 & 0.665820 & 2316 & 2422 & 2319 & 2419 & 0.665816 \\
12 & 2145 & 2244 & 0.665820 & 2319 & 2425 & 2322 & 2422 & 0.665816 \\
Final& 2140(24) & 2239(25) & 0.665821(2) & 2314(25) & 2420(26) & 2317(25) & 2417(26) & 0.665817(4) \\
\\
\multicolumn{9}{c}{$4d^{10}5p\ ^2P_{3/2}$}\\
5 & 546 & 571 & 1.334062 & 644 & 674 & 650 & 680 & 1.334056 \\
6 & 699 & 732 & 1.334057 & 787 & 824 & 791 & 827 & 1.334052 \\
7 & 699 & 731 & 1.334060 & 768 & 803 & 772 & 807 & 1.334056 \\
8 & 727 & 761 & 1.334060 & 794 & 830 & 797 & 834 & 1.334056 \\
9 & 722 & 755 & 1.334062 & 786 & 822 & 789 & 826 & 1.334058 \\
10 & 729 & 763 & 1.334060 & 789 & 828 & 793 & 832 & 1.334055 \\
11 & 728 & 762 & 1.334059 & 789 & 824 & 792 & 828 & 1.334057 \\
12 & 729 & 762 & 1.334060 & 789 & 825 & 793 & 829 & 1.334056 \\
Final& 727(8) & 760(8) & 1.334061(2) & 789(8) & 823(9) & 792(8) & 830(9) & 1.334056(3) \\
\hline\hline
\end{tabular}
\end{table*}

In our second calculation, based on the above investigations, we included CV and CC correlations for the $4d$, $4p$, $4s$, $3d$, and $3p$ electrons, as well as the CV correlation down to the $1s$ subshell by allowing restricted SD excitations to $n\leq12, l\leq6$ (AS$_{12}$), in the RSCF calculation. RCI calculations were also performed to include the Breit and QED effects.
Note that this calculation also started from the MCDHF calculation for the $4d^{10}5s$ and $4d^{10}5p$ configurations and hence is labelled MCDHF2+RCI.

In analyzing the wavefunction compositions from the MCDHF2+RCI calculation, we noticed strong correlations between $4d^{10}5s$, $4d^84f^25s$, and $4d^94f5p$ configurations, and between $4d^{10}5p$, $4d^84f^25p$, and $4d^85p5d^2$ configurations. Therefore, instead of starting from the DHF calculation where only $4d^{10}5s$ and $4d^{10}5p$ were included in the CSF list, we allowed the $4d$ and $5s/5p$ electrons to be SD-excited to \{$5s,5p,5d,4f$\} to generate the CSF list as a starting point of our third calculation approach. In this way, the spectroscopic orbitals together with the $5d$ and $4f$ orbitals are optimized together, and the correlation effect between the essential CSFs is included in the beginning. The CV and CC correlation effects are included by systematically increasing the virtual excitations to AS$_{12}$; this calculation method is labelled MCDHF3+RCI.

For $4d^{10}5s\ ^2S_{1/2}$, because there are no other levels with which to have strong hyperfine interactions, we only included the diagonal contributions to its HFS. The calculated HFSs and $g_J$ factors with an increasing AS size from MCDHF2+RCI and MCDHF3+RCI calculations are listed in Table~\ref{tab_S1}. We find some fluctuations in our calculated HFSs with increasing AS size, but the values from the last few ASs generally tend to some specific value. We, therefore, took the average of the last three values (AS$_{10}$, AS$_{11}$, and AS$_{12}$) as our final calculated result, with the maximum difference between them taken as the calculation error. Although the final splitting for $^{113}$Cd$^+$ from the MCDHF2+RCI calculation (i.e., 14563(7)~MHz) is much smaller than the experimental measurement (i.e., 15199~MHz), the MCDHF3+RCI calculation, (15206(9)~MHz) shows a significant improvement with the experimental value being within the estimated uncertainty of the latter calculation. Following a similar method, the $g_J$ factors of $2S_{1/2}$ from MCDHF2+RCI and MCDHF3+RCI calculations were 2.002262(2) and 2.002257(1), respectively. The HFSs and $g_J$ factors for $4d^{10}5s\ ^2P_{1/2,3/2}$ with an increasing AS are listed in Table~\ref{tab_P1}. Following the same method as used in determining our final calculation results and their uncertainties, the HFSs for the $^2P_{3/2}$ level of $^{111}$Cd$^{+}$/$^{113}$Cd$^{+}$ from MCDHF2+RCI and MCDHF3+RCI calculations when not including the off-diagonal contributions are 727(8)/760(8)~MHz and 789(8)/823(9)~MHz, respectively. With off-diagonal contributions included, the MCDHF3+RCI results increase to 792(8)/830(9)~MHz.

\section{Results and discussions}
The measured HFSs for the $5p~^2P_{3/2}$ level and the calculated HFSs and Land\'{e} $g_J$ factors for the $5s~^2S_{1/2}$ and $5p~^2P_{1/3,3/2}$ levels in this work are listed in Table~\ref{tab:1}; other experimental and calculated results are also listed for comparison. For the HFSs, our group's previous high-accuracy measurements for the $^{111,113}$Cd$^+$ ground state provided an excellent benchmark for the atomic structure calculation of Cd$^+$. The present HFSs for the $5s~^2S_{1/2}$ state calculated using the MCDHF method show stronger agreement with our previous experimental results than those of previous theoretical results \cite{dixit2008ab, li2018relativistic}. The present measured HFSs for the $5p~^2P_{3/2}$ level is also in agreement with the present theoretical results. The cross-checking between experiment and theory ensures the reliability of the Cd$^+$ $5p~^2P_{3/2}$ HFSs determined in this work. We recommend the adoption of 794.6(3.6) and 835.5(2.9) as the blue-shifted frequencies for optical pumping in the microwave-frequency standard based on $^{111/113}$Cd$^+$.

Regarding the Land\'{e} $g_J$ factors, there are no experimental results for Cd$^+$. Accurate calculations of Land\'{e} $g_J$ factors has proven complicated even for alkali atoms and alkali-like ions because they are sensitive to electron correlations. Those calculated in this work using the MCDHF method show strong deviations from previous RCC results \cite{han2019roles}. The ground state Land\'{e} $g_J$ factor calculated in this work ($2.002257(1)$) agrees with the recent result calculated using the $\Lambda$-RCC theory ($2.002291(4)$) \cite{Yu2020} to the fourth decimal place, although there is no overlap within their margins of uncertainty. To our knowledge, there also exists a significant difference in results between the $\Lambda$-RCC calculations with the configuration interaction and the many-body perturbation (CI+MBPT) calculations in Yb$^+$ ground-state Land\'{e} $g_J$ factor \cite{gossel2013calculation, Yu2020}. Comparing the results of the same physical quantity from different calculation methods is also of great significance for developing atomic structure calculation models and understanding the role of electronic correlation effects. Therefore, we encourage more experimental and theoretical research on the Land\'{e} $g_J$ factors of Cd$^+$.

For precaution, we recommend the value 2.00226(4) for the Cd$^+$ ground state Land\'{e} $g_J$ factor in the evaluation of the SOZS of the microwave frequency standard of trapped Cd$^+$ ions. The SOZS can be estimated using the Breit--Rabi formula,
\begin{equation}
\Delta \nu^{(2)}_{Zeem}(B)=-\frac{[g_j-g_I]^2\mu_B^2B^2}{2h^2A_{hf}}, \label{2ndZeem}
\end{equation}
for which $B\sim8000$ nT for the Cd$^+$ microwave frequency standard during actual operations. Thus, the fractional frequency shifts incurred when using the value of $g_J=2.00226(4)$ is $4.4\times10^{-15}$. The fractional frequency shifts produced by this $g_J$ factor for the Cd$^+$ ground-state can meet current accuracy requirements for the best Cd$^+$ microwave frequency standard ($1.8\times10^{-14}$). However, for further developments of this standard, the ground state $g_J$ factor of Cd$^+$ also needs to be determined more accurately.

\begin{table}
\caption{Measured HFSs for $5p~^2P_{3/2}$ and the calculated HFSs and Land\'{e} $g_J$ factors of $5s~^2S_{1/2}$ and $5p~^2P_{1/3,3/2}$ of this work. Results of the Cd$^+$ HFSs and Land\'{e} $g_J$ factors from other works are also listed for comparison.} \label{tab:1}
{\setlength{\tabcolsep}{2pt}
\begin{tabular}{llllllll}\hline\hline
$^{111}$HFS & $^{113}$HFS & $g_J$ & Method \\ \hline
\multicolumn{4}{c}{ $5s~^2S_{1/2}$ } & \\
14530.507 & 15199.863 & & Exp. \cite{zhang2012} \\
14536(9) & 15206(9) & 2.002257(1) & MCDHF (This work) \\
14478(175) & 15146(183) & & RCC \cite{li2018relativistic} \\
& 15280 & & RCC \cite{dixit2008ab} \\
& & 2.00286(53) & RCC \cite{han2019roles} \\
& & 2.002291(4) & $\Lambda$-RCC \cite{yu2020ground} \\ \\
\multicolumn{4}{c}{ $5p~^2P_{3/2}$ } & \\
794.6(3.6) & 835.5(2.9) & & Exp. (This work) \\
& 800 & & Exp. \cite{tanaka1996determination} \\
792(8) & 830(9) & 1.334056(3) & MCDHF (This work) \\
794(12) & 832(12) & & RCC \cite{li2018relativistic} \\
& 812.04 & & RCC \cite{dixit2008ab} \\
& & 1.33515(43) & RCC \cite{han2019roles} \\ \\
\multicolumn{4}{c}{ $5p~^2P_{1/2}$ } & \\
2317(25) & 2417(26) & 0.665817(4) & MCDHF (This work) \\
2333(31) & 2441(33) & & RCC \cite{li2018relativistic} \\
& 2430 & & RCC \cite{dixit2008ab} \\
& & 0.66747(83) & RCC \cite{han2019roles} \\
\hline\hline
\end{tabular}}
\end{table}

\section{Conclusion}
We reported on the determination of HFSs and Land\'{e} $g_J$ factors for the $5s~^2S_{1/2}$ and $5p~^2P_{1/2,3/2}$ levels of $^{111,113}$Cd$^+$. The HFSs of the $5p~^2P_{3/2}$ level was measured using the laser-induced-fluorescence technique. The Cd$^+$ ions were co-trapped with Ca$^+$ ions in the same linear ion trap and sympathetically cooled through the Coulomb interaction with laser-cooled Ca$^+$ ions. Furthermore, the HFSs and Land\'{e} $g_J$ factors for both levels of interest were calculated using the MCDHF calculation. Three computational strategies were followed to account for the electronic correlation effects more comprehensively. The final calculated HFSs were in perfect agreement with the measured HFSs of this work and our previous work, which from cross-checks, demonstrated the reliability of the calculations and the experiments. The HFSs and Land\'{e} $g_J$ factors determined in this work can further improve the efficiency of the optical pumping procedure and the accuracy of the second-order Zeeman correction, and the stability and accuracy of the microwave frequency standard based on trapped Cd$^+$ ions.

\section*{Acknowledgements}
We thank Z. M. Tang for the helpful discussions. This work is supported by the National Key R\&D Program of China (No. 2021YFA1400243), National Natural Science Foundation of China (Nos. 91436210, 12074081, 12104095).

\bibliography{apssamp}
\end{document}